\documentclass{iopart}

\usepackage{iopams}
\usepackage{graphicx}  
\usepackage{subfigure}
\usepackage{sidecap}
\usepackage{setstack}

\begin{document}

\title{Triplet superconductivity and proximity effect induced by Bloch
  and N\'{e}el domain walls}

\author{Daniel Fritsch\footnote{Present address: Department of
    Chemistry, University of Bath, Claverton Down, Bath, BA2 7AY, UK.}
  and James F. Annett}

\address{H. H. Wills Physics Laboratory, School of Physics, University
  of Bristol, Bristol BS8 1TL, UK}

\ead{d.fritsch@bath.ac.uk}

\begin{abstract}
Noncollinear magnetic interfaces introduced in superconductor
(SC)/ferromagnet/SC heterostructures allow for spin-flipping processes
and are able to generate equal-spin spin-triplet pairing correlations
within the ferromagnetic region. This leads to the occurrence of the
so-called long-range proximity effect. Particular examples of
noncollinear magnetic interfaces include Bloch and N\'{e}el domain
walls. Here, we present results for heterostructures containing Bloch
and N\'{e}el domain walls based on self-consistent solutions of the
spin-dependent Bogoliubov$-$de Gennes equations in the clean limit. In
particular, we investigate the thickness dependence of Bloch and
N\'{e}el domain walls on induced spin-triplet pairing correlations and
compare with other experimental and theoretical results, including
conical magnetic layers as noncollinear magnetic interfaces. It is
shown that both, Bloch and N\'{e}el domain walls lead to the
generation of unequal-spin spin-triplet pairing correlations of
similar strength as for conical magnetic layers. However, for the
particular heterostructure geometries investigated, only Bloch domain
walls lead to the generation of equal-spin spin-triplet pairing
correlations. They are stronger than those generated by an equivalent
thickness of conical magnetic layers. In order for N\'{e}el domain
walls to induce equal-spin spin-triplet pairing correlations, they
have to be oriented such that the noncollinearity appears within the
plane parallel to the interface region.
\end{abstract}

\submitto{\SUST}

\section{Introduction}
\label{Introduction}

Bringing a superconductor (SC) and a ferromagnetic (FM) material in
close proximity to each other results in drastic changes for the
spin-singlet Cooper pairs. While the Pauli principle requires the
spins to orient parallel in the FM region, the spin-singlet Cooper
pair electrons tend to align antiparallel in the SC region. This leads
to the following consequences: i) Different Fermi velocities in the
spin-up and spin-down channel of the electrons due to the exchange
field lead to a centre of mass modulation and an oscillating behaviour
of the superconducting correlations within the FM
region~\cite{Buzdin_JETPLett35_178,Buzdin_JETPLett53_321,Demler_PRB55_15174},
namely the FFLO
oscillations~\cite{Fulde_PhysRev135_A550,Larkin_JETP20_762}. Unequal-spin
spin-triplet correlations ($S_{z} = 0$) are also generated. Both of
these correlations are oscillating and suppressed in the FM region,
and are essentially short-range. ii) Following a theoretical
suggestion by Bergeret \textit{et al.}~\cite{Bergeret_PRL86_4096} it
should also be possible to induce equal-spin spin-triplet Cooper pairs
which are compatible with and unaffected by the FM exchange field,
thus leading to much larger penetration depths. This phenomenon
requires spin-flip processes at the interface and is called the
long-range proximity effect.

Since this first theoretical prediction equal-spin spin-triplet
pairing correlations have attracted a lot of attention and are
reviewed by Buzdin~\cite{Buzdin_RMP77_935}, Bergeret \textit{et
  al.}~\cite{Bergeret_RMP77_1321}, and Tanaka \textit{et
  al.}~\cite{Tanaka_JPSJ81_011013}. Several multilayer setups have
been suggested experimentally and theoretically to shed some light on
the specific interface properties necessary to generate equal-spin
spin-triplet pairing correlations within such structures. These setups
typically involve noncollinear magnetic structures within the
interface region as provided by different FM
layers~\cite{Klose_PRL108_127002,Gingrich_PRB86_224506}, helical (or
conical) magnetic
layers~\cite{Volkov_PRB73_104412,Robinson_Science329_59,Alidoust_PRB81_014512,Alidoust_PRB82_224504,Fritsch_NJP16_055005},
or through specific interface
potentials~\cite{Bozovic_EPL70_513,Linder_PRL102_107008,Linder_PRB81_214504,Terrade_PRB88_054516}. Moreover,
also the effects of
Bloch~\cite{Alidoust_PRB81_014512,Alidoust_PRB82_224504,Bergeret_PRL110_117003}
and
N\'{e}el~\cite{Alidoust_PRB81_014512,Alidoust_PRB82_224504,Bergeret_PRL110_117003,Fominov_PRB75_104509}
domain walls, as well as interfacial spin-orbit
coupling~\cite{Lv_EurPhysJB83_493,Yang_SSTech22_055012} as source for
generating equal-spin spin-triplet pairing correlations have been
investigated.

On the theoretical side, equal-spin spin-triplet pairing correlations
have been investigated using Green's function techniques based on
solutions of the Usadel
equations~\cite{Volkov_PRB73_104412,Alidoust_PRB81_014512,Alidoust_PRB82_224504,Bergeret_PRL110_117003},
and self-consistent solutions of the Bogoliubov$-$de Gennes (BdG)
equations~\cite{Fritsch_NJP16_055005,Bozovic_EPL70_513}. While the
Green's function techniques allow for an easier inclusion of
scattering effects describing a shorter electron mean free path in the
diffusive limit, these effects can in principle be incorporated into
the BdG equations as well via additional impurity scattering
potentials. However, in the present work we are solely concerned with
the special case of odd-frequency triplet pairing correlations, which
are not affected by nonmagnetic
impurities~\cite{Bergeret_RMP77_1321}. The inclusion of additional
impurity scattering potentials would require further investigations.

The main aim of the paper is to assess the effectiveness of Bloch and
N\'{e}el domain walls in comparison to conical magnetic layers to
generate long-range proximity effects. This is motivated by the
potential use of magnetic domain walls incorporated in experimental
heterostructures to allow for a switching of the generated long-range
spin-triplet Josephson current. Based on an established
heterostructure setup successfully used in previous experimental and
theoretical works we introduce Bloch and N\'{e}el domain walls in the
interface regions and investigate its suitability to generate
equal-spin spin-triplet pairing correlations. These results will be
compared to the corresponding results with interface regions
containing conical magnetic layers of varying thicknesses.

This paper is organised as follow. \Sref{Sec2} gives an overview over
the theoretical methods. This includes a description of the
microscopic BdG equations in \sref{Sec2_1}, and the definition of the
spin-triplet pairing correlations in \sref{Sec2_2}. The results for
heterostructures containing Bloch and N\'{e}el domain walls of varying
thicknesses $n_{\rm Bloch}$ and $n_{\rm N\acute{e}el}$ are presented
in \sref{Sec3_1} and \sref{Sec3_2},
respectively. \Sref{SummaryAndOutlook} provides a summary and an
outlook.

\section{Theoretical background}
\label{Sec2}

\subsection{BdG equations and heterostructure setup}
\label{Sec2_1}

The results presented in this paper are obtained by means of the
microscopic BdG equations, which have been solved self-consistently in
the clean limit. In the most general spin-dependent case and
incorporating an expression for arbitrary exchange fields {\bf h}
needed later to describe the influence of Bloch and N\'{e}el domain
walls, the BdG equations
read~\cite{Fritsch_NJP16_055005,Annett_book,KettersonSong_Superconductivity}
\begin{eqnarray}
  \fl \label{EqBdGGeneral} \left(
  \begin{array}{cccc}
    {\cal H}_{0} - h_{z} & -h_{x} + i h_{y} & \Delta_{\uparrow
      \uparrow} & \Delta_{\uparrow \downarrow} \\ - h_{x} - i h_{y} &
    {\cal H}_{0} + h_{z} & \Delta_{\downarrow \uparrow} &
    \Delta_{\downarrow \downarrow} \\ \Delta_{\uparrow \uparrow}^{*} &
    \Delta_{\downarrow \uparrow}^{*} & -{\cal H}_{0} + h_{z} & h_{x} +
    i h_{y} \\ \Delta_{\uparrow \downarrow}^{*} & \Delta_{\downarrow
      \downarrow}^{*} & h_{x} - i h_{y} & -{\cal H}_{0} - h_{z}
  \end{array}
  \right) \left(
  \begin{array}{c}
    u_{n\uparrow} \\ u_{n\downarrow} \\ v_{n\uparrow}
    \\ v_{n\downarrow}
  \end{array}
  \right) =\varepsilon_{n} \left(
  \begin{array}{c}
    u_{n\uparrow} \\ u_{n\downarrow} \\ v_{n\uparrow}
    \\ v_{n\downarrow}
  \end{array}
  \right) \,,
\end{eqnarray}
where $\varepsilon_{n}$, $u_{n\sigma}$, and $v_{n\sigma}$ denote
eigenvalues, and quasiparticle and quasihole amplitudes for spin
$\sigma$, respectively. Following simplifications introduced
earlier~\cite{Fritsch_NJP16_055005,Sipr_JPCM7_5239,Covaci_PRB73_014503}
the tight-binding Hamiltonian ${\cal H}_{0}$ in \eref{EqBdGGeneral}
becomes effectively one-dimensional and reads
\begin{equation}
  \label{EqTBHamiltonianLinear}
        {\cal H}_{0} = -t \sum_{n}{ \left( c_{n}^{\dagger}c_{n+1} +
          c_{n+1}^{\dagger}c_{n} \right) } + \sum_{n}{ \left(
          \varepsilon_{n} - \mu \right) c_{n}^{\dagger} c_{n} } \,.
\end{equation}
Therein, $c_{n}^{\dagger}$ and $c_{n}$ denote electronic creation and
destruction operators at multilayer index $n$, respectively. In order
to compare the present results with those obtained
previously~\cite{Fritsch_NJP16_055005,Fritsch_JPCM26_274212,Fritsch_PhilMag95_441}
we choose the next-nearest neighbour hopping parameter $t = 1$ and set
the energy scales via the chemical potential (Fermi energy) $\mu = 0$.

Balian and Werthamer~\cite{Balian_PhysRev131_1553,Sigrist_RMP63_239}
introduced a way to rewrite the general form of the pairing matrix
appearing in \eref{EqBdGGeneral} as
\begin{equation}
  \fl \label{EqBalianWerthamer} \left( \begin{array}{cc}
    \Delta_{\uparrow \uparrow} & \Delta_{\uparrow \downarrow}
    \\ \Delta_{\downarrow \uparrow} & \Delta_{\downarrow
      \downarrow} \end{array} \right) = \left( \Delta + {\hat{\bf
      \sigma}} \cdot {\bf d}\right) i \hat{\sigma_{2}} =
  \left( \begin{array}{cc} -d_{x} + i d_{y} & \Delta + d_{z} \\ -
    \Delta + d_{z} & d_{x} + i d_{y} \end{array} \right) \,.
\end{equation}
Here, $\hat{\cdots}$ denotes a $2 \times 2$ matrix. Making additional
use of the Pauli matrices $\hat{\bf \sigma}$, the superconducting
order parameter is now written as a singlet (scalar) part $\Delta$ and
a triplet (vector) part {\bf d}, respectively. This work is restricted
to $s$-wave superconductors, and the pairing potential entering
\eref{EqBdGGeneral} simplifies to a scalar quantity $\Delta$, which
fulfills the self-consistency condition
\begin{equation}
  \label{EqDeltaSelfConsistency}
  \Delta({\bf r}) = \frac{g({\bf r})}{2} \sum_{n}{ \bigl(
    u_{n\uparrow}({\bf r})v_{n\downarrow}^{*}({\bf r})
    [1-f(\varepsilon_{n})] + u_{n\downarrow}({\bf
      r})v_{n\uparrow}^{*}({\bf r}) f(\varepsilon_{n}) \bigr)} \,.
\end{equation}
The sum is evaluated only over positive eigenvalues $\varepsilon_{n}$
and $f(\varepsilon_{n})$ denotes the Fermi distribution function
evaluated as a step function for zero temperature.

\begin{figure}
  \centering
  \includegraphics[width=0.95\textwidth,clip]{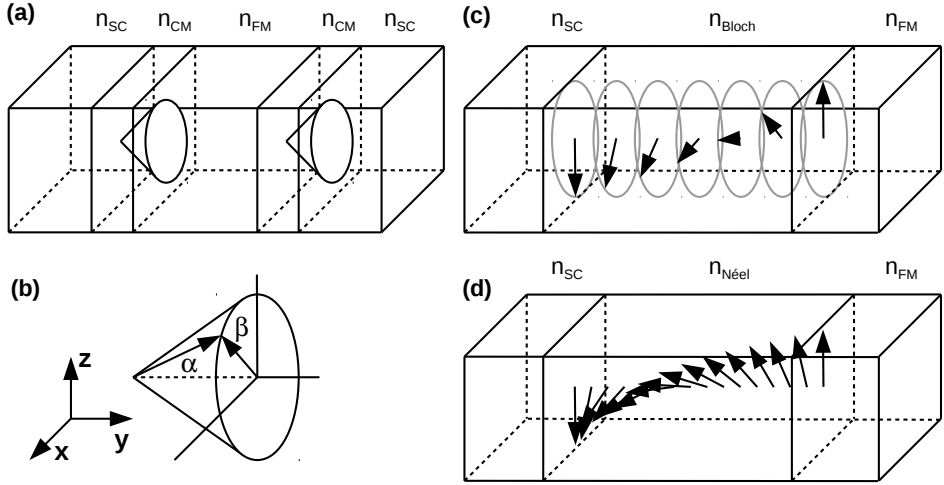}
  \caption{\label{Fig1} Structural setup of the heterostructures
    investigated in this work: (a) General setup including $n_{\rm
      SC}$ layers of $s$-wave superconductor to the left, $n_{\rm CM}$
    layers of a conical magnet, $n_{\rm FM}$ layers of ferromagnetic
    middle layer, and $n_{\rm CM}$ conical magnet and $n_{\rm SC}$
    $s$-wave superconductor layers to the right. (b) Employed
    cordinate system and definition of the opening angle $\alpha$ and
    the turning (or pitch) angle $\beta$ of the conical magnetic
    structure. (c) Orientation of the Bloch domain wall of $n_{\rm
      Bloch}$ layers between the left $s$-wave superconductor and the
    middle ferromagnetic region. (d) Orientation of the N\'{e}el
    domain wall of $n_{\rm N\acute{e}el}$ layers between the left
    $s$-wave superconductor and the middle ferromagnetic region. Note
    that (c) and (d) only show the left interface region for clarity.}
\end{figure}

The heterostructure is set up as generally shown in \fref{Fig1}(a)
with a lattice constant $a = 1$ and a strength of the exchange field
$h_{0} = 0.1$. The effective superconducting coupling parameter
$g({\bf r})$ equals $1$ in the $n_{\rm SC}=250$ left and right layers
of spin-singlet $s$-wave superconductor and vanishes
elsewhere. Between the $s$-wave superconductors and $n_{\rm FM}=100$
layers of the ferromagnetic middle layer we introduce conical magnetic
layers (\fref{Fig1}(a)), and Bloch (\fref{Fig1}(c)) and N\'{e}el
(\fref{Fig1}(d)) domain walls of varying thicknesses $n_{\rm CM}$,
$n_{\rm Bloch}$, and $n_{\rm N\acute{e}el}$, respectively. The conical
magnetic layers are arranged as in our previous
works~\cite{Fritsch_NJP16_055005,Fritsch_PhilMag95_441} and are chosen
to represent the conical magnet Holmium. The maximum number of conical
magnetic layers considered here correspond to one full turn of the
magnetisation around the cone. The Holmium opening angle $\alpha =
80^{\circ}$ and turning (or pitch) angle $\beta = 30^{\circ}$ are
measured from the positive $y$ axis towards the positive $z$ axis and
from the positive $z$ axis towards the positive $x$ axis, respectively
(\fref{Fig1}(b)). The Bloch and N\'{e}el domain walls are arranged
such that there is a ferromagnetic coupling to the middle
ferromagnetic region of the heterostructure and that a C$_{2}$
symmetry along the central $z$ axis is retained.

\subsection{(Triplet) pairing correlations}
\label{Sec2_2}

The specific superconducting pairing correlation between spins
$\alpha$ and $\beta$ we are interested in here is evaluated as on-site
interaction for times $t = \tau$ and $t' = 0$ as
\begin{equation}
  \label{EqPairingCorrelationGeneral}
  f_{\alpha \beta}({\bf r}, \tau, 0) = \frac{1}{2}\bigl<
  \hat{\Psi}_{\alpha}({\bf r},\tau) \hat{\Psi}_{\beta}({\bf r},0)
  \bigr> \,.
\end{equation}
Therein, $\hat{\Psi}_{\sigma}({\bf r},\tau)$ denotes the many-body
field operator for spin $\sigma$ at time $\tau$, and time-dependence
is introduced through the Heisenberg equation of motion. The
particular pairing correlation of \eref{EqPairingCorrelationGeneral}
is local in space and leads to vanishing triplet contributions for
$\tau = 0$, in accordance with the Pauli
principle~\cite{Halterman_PRL99_127002}. However, finite values of
$\tau$ give rise to nonvanishing pairing correlations, being an
example of odd-frequency triplet
pairing~\cite{Bergeret_RMP77_1321}. Substituting the field operators
valid for our setup and phase convention the spin-dependent triplet
pairing correlations
read~\cite{Fritsch_JPCM26_274212,Fritsch_PhilMag95_441}
\begin{equation}
  \label{EqTripletPairingCorrelations}
  \fl \eqalign { f_{0}(y, \tau) = \frac{1}{2} \bigl( f_{\uparrow
      \downarrow}(y, \tau) + f_{\downarrow \uparrow}(y, \tau) \bigr) =
    \frac{1}{2} \sum_{n} {\bigl( u_{n\uparrow}(y)
      v_{n\downarrow}^{*}(y) + u_{n\downarrow}(y) v_{n\uparrow}^{*}(y)
      \bigr) \zeta_{n}(\tau)} \cr f_{1}(y, \tau) = \frac{1}{2} \bigl(
    f_{\uparrow \uparrow}(y, \tau) - f_{\downarrow \downarrow}(y,
    \tau) \bigr) = \frac{1}{2} \sum_{n} {\bigl( u_{n\uparrow}(y)
      v_{n\uparrow}^{*}(y) - u_{n\downarrow}(y) v_{n\downarrow}^{*}(y)
      \bigr) \zeta_{n}(\tau)} } \,,
\end{equation}
depending on position $y$ within the heterostructure and time
parameter $\tau$ (set to $\tau = 10$ in the present work), and with
$\zeta_{n}(\tau)$ given by
\begin{equation}
  \label{EqTau}
  \zeta_{n}(\tau) = \cos (\varepsilon_{n} \tau) - i \sin
  (\varepsilon_{n} \tau) \bigl( 1 - 2 f(\varepsilon_{n}) \bigr) \,.
\end{equation}
Based on the rewritten form of the pairing matrix in
\eref{EqBalianWerthamer} and defining $\hat{\Delta}$ as the triplet
pairing matrix for an ordinary spin-triplet superconductor, the
product ${\hat \Delta}{\hat \Delta}^{\dagger}$ can be written as
\begin{equation}
  \label{EqDDdagger}
  {\hat \Delta}{\hat \Delta}^{\dagger} = |{\bf d}|^{2} {\hat
    \sigma}_{0} + i \left( {\bf d}\times{\bf d}^{*} \right) {\hat {\bf
      \sigma}} \,.
\end{equation}
Therein, the magnitude of the ${\bf d}$-vector denotes the gap
function independently from the underlying coordinate system, and
i${\bf d}\times{\bf d}^{*}$ is a measure of the spin magnetic moment,
respectively. In the present work there is no pairing interaction
considered in the triplet channel, i.e., $g({\bf r}) = 1$ in
\eref{EqDeltaSelfConsistency}. However, triplet pairing correlations
are induced which are given by the so-called triplet pair correlation
function matrix instead, which can be written similarly to the pairing
matrix $\hat{\Delta}$
as~\cite{Kawabata_JPSJ82_124702,Fritsch_PhilMag95_441}
\begin{equation}
  \label{EqPairFunctionMatrix} {\hat f} = \left( {\hat{\bf
      \sigma}} \cdot {\bf f}\right) i {\hat \sigma}_{2} =
  \left( \begin{array}{cc} -f_{x} + i f_{y} & f_{z} \\ f_{z} & f_{x} +
    i f_{y} \end{array} \right) \,.
\end{equation}
With \eref{EqPairFunctionMatrix} the analogue to \eref{EqDDdagger} can
be written as
\begin{equation}
  \label{EqFFdagger}
        {\hat f}{\hat f}^{\dagger} = |{\bf f}|^{2} {\hat \sigma}_{0} +
        i \left( {\bf f}\times{\bf f}^{*} \right) {\hat {\bf \sigma}}
        \,.
\end{equation}
Analogously to $|{\bf d}|$ and ${\bf d}\times{\bf d}^{*}$ we find
$|{\bf f}|$ and ${\bf f}\times{\bf f}^{*}$. Again, they can be
rewritten in terms of the ${\bf f}$-vector components which depend on
the spin-dependent triplet pairing correlations as
\begin{equation}
  \label{EqDVector}
  \eqalign { f_{x} = \frac{1}{2} \left( -f_{\uparrow \uparrow} +
    f_{\downarrow \downarrow} \right) \cr f_{y} = - \frac{i}{2} \left(
    f_{\uparrow \uparrow} + f_{\downarrow \downarrow} \right) \cr
    f_{z} = \frac{1}{2} \left( f_{\uparrow \downarrow} + f_{\downarrow
      \uparrow} \right) } \,.
\end{equation}

\section{Results and Discussion}
\label{Sec3}

\subsection{Heterostructure including Bloch domain walls}
\label{Sec3_1}

The results presented in this section are obtained for
heterostructures containing Bloch domain walls of increasing thickness
$n_{\rm Bloch}$ to either side of the ferromagnetic middle layer. The
Bloch domain walls are oriented such that the magnetic moments at the
interface align parallel to the ferromagnetic middle layers
(\fref{Fig1}(c)) and that a C$_{2}$ symmetry along the central $z$
axis is retained. Recently, heterostructures containing conical
magnetic layers of different opening and turning angles leading to
different symmetries have been
investigated~\cite{Fritsch_NJP16_055005,Fritsch_PhilMag95_441} and
could be explained with the properties of the ${\bf f}$-vector
introduced in \eref{EqTripletPairingCorrelations}.
\begin{figure}
  \centering
  \includegraphics[width=0.95\textwidth,clip]{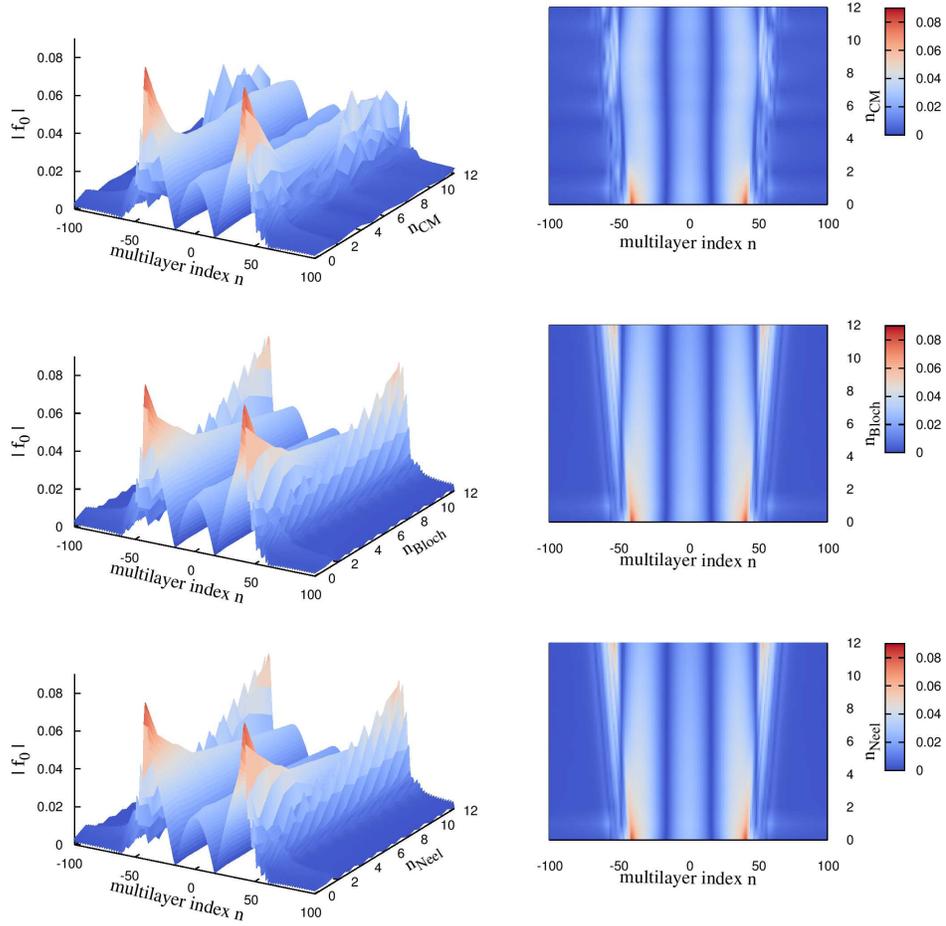}
  \caption{\label{Fig2} Magnitude of the unequal-spin spin-triplet
    pairing correlations ${\bf f}_{0}$ for heterostructures containing
    conical magnetic layers $n_{\rm CM}$ (upper panels), Bloch domain
    walls $n_{\rm Bloch}$ (middle panels), and N\'{e}el domain walls
    $n_{\rm N\acute{e}el}$ (lower panels), respectively. The left and
    right panels both show the full data sets and the right panels
    show a top view of the data.}
\end{figure}
The fundamental pairing correlations evaluated according to
\eref{EqTripletPairingCorrelations} are shown in \fref{Fig2} and
\fref{Fig3} for heterostructures containing conical magnetic layers
$n_{\rm CM}$, Bloch domain walls $n_{\rm Bloch}$, and N\'{e}el domain
walls $n_{\rm N\acute{e}el}$, respectively. In particular, the middle
panels of \fref{Fig2} show the magnitude of the unequal-spin
spin-triplet pairing correlations ${\bf f}_{0}$ for increasing
thickness of the Bloch domain wall $n_{\rm Bloch}$ in comparison to
conical magnetic layers $n_{\rm CM}$ (upper panels) and N\'{e}el
domain wall $n_{\rm N\acute{e}el}$ (lower panels), respectively. The
full data sets of the left panels are shown as top views in the
respective right panels.

Looking at the oscillations within the ferromagnetic middle region of
the heterostructure, one notices strong oscillations of the
unequal-spin spin-triplet pairing correlations ${\bf f}_{0}$. This is
very similar to heterostructures containing conical magnetic layers
(shown in the upper panels). These are essentially the FFLO
oscillations discussed above. The magnitude is almost equally strong
for thin Bloch domain walls, but decays to smaller values for
increasing Bloch domain wall thickness compared to the conical
magnetic layers. However, within the interface region the magnitude of
${\bf f}_{0}$ is increasing linearly with Bloch domain wall thickness
and clearly surpasses the strength in case of the conical magnetic
layers.
\begin{figure}
  \centering
  \includegraphics[width=0.95\textwidth,clip]{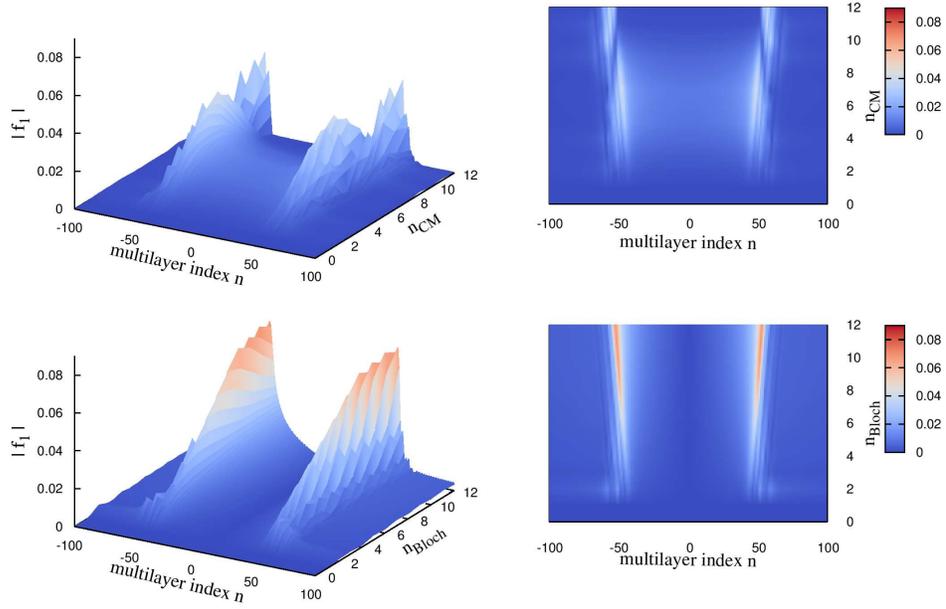}
  \caption{\label{Fig3} Magnitude of the equal-spin spin-triplet
    pairing correlations ${\bf f}_{1}$ for heterostructures containing
    conical magnetic layers $n_{\rm CM}$ (upper panels), and Bloch
    domain walls $n_{\rm Bloch}$ (lower panels), respectively. Note
    that the ${\bf f}_{1}$ contribution from heterostructures
    containing N\'{e}el domain walls is zero for the chosen geometry
    (as discussed in the text). The left and right panels both show
    the full data sets and the right panels show a top view of the
    data.}
\end{figure}
Looking now at \fref{Fig3} showing the respective results for the
magnitude of the equal-spin spin-triplet pairing correlations ${\bf
  f}_{1}$ for heterostructures containing conical magnetic layers
$n_{\rm CM}$ (upper panels) and Bloch domain walls $n_{\rm Bloch}$
(lower panels) one immediately notices larger differences. The
magnitude of ${\bf f}_{1}$ rises linearly and is similarly strong for
smaller values of Bloch domain wall thickness $n_{\rm Bloch}$ compared
to the same thickness of conical magnetic layers $n_{\rm CM}$. In
contrary to the conical magnetic layers, for which ${\bf f}_{1}$
starts to oscillate and decay with a further increase of layer
thickness again, the equal-spin spin-triplet pairing correlations
${\bf f}_{1}$ for thicker Bloch domain walls increase further and
reach saturation for the maximum thickness considered here. Moreover,
the overall magnitude of ${\bf f}_{1}$ for heterostructures containing
Bloch domain walls clearly exceeds those for heterostructures
containing conical magnetic layers.
\begin{figure}
  \centering
  \includegraphics[width=0.95\textwidth,clip]{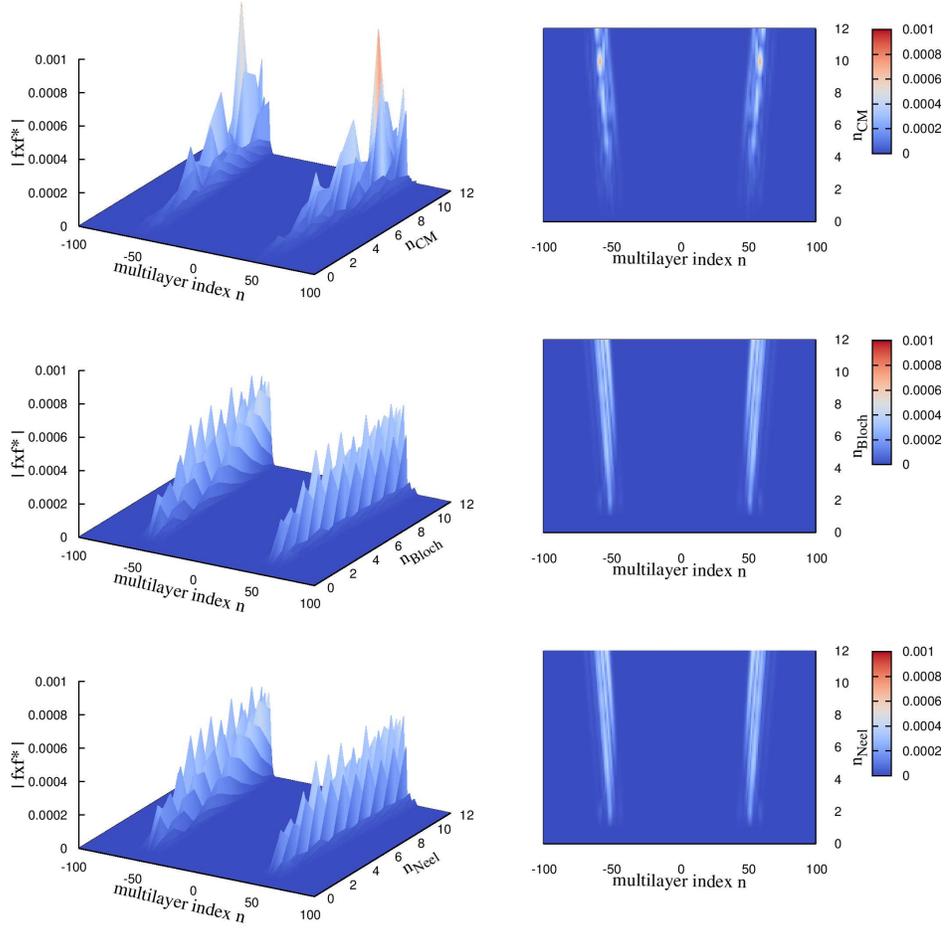}
  \caption{\label{Fig4} Magnitude of ${\bf f} \times {\bf f}^{*}$ for
    heterostructures containing conical magnetic layers $n_{\rm CM}$
    (upper panels), Bloch domain walls $n_{\rm Bloch}$ (middle
    panels), and N\'{e}el domain walls $n_{\rm N\acute{e}el}$ (lower
    panels), respectively. The left and right panels both show the
    full data sets and the right panels show a top view of the data.}
\end{figure}
Finally, a measure for the induced spin magnetic moment provided by
the quantity ${\bf f}\times{\bf f}^{*}$ of \eref{EqFFdagger} and
\eref{EqDVector} is shown in \fref{Fig4}, where again the
contributions from heterostructures containing conical magnetic layers
$n_{\rm CM}$, and Bloch $n_{\rm Bloch}$ and N\'{e}el $n_{\rm
  N\acute{e}el}$ domain walls are shown in the upper, middle, and
lower panels, respectively. One immediately notices that spin
magnetisation only occurs in the respective interface regions and that
no magnetisation leaks into the SC regions of the
heterostructures. The overall rise in spin magnetisation is stronger
for the Bloch domain walls and saturates earlier than in the conical
magnetic layers. The latter also show stronger oscillations depending
on the thickness $n_{\rm CM}$ compared to the Bloch domain wall
thickness $n_{\rm Bloch}$.

Summarising this we find a similar influence on ${\bf f}_{0}$ for
heterostructures containing conical magnetic layers and Bloch domain
walls within the ferromagnetic middle layer, and a stronger influence
within the Bloch domain walls for increasing thickness $n_{\rm
  Bloch}$. The magnitude of ${\bf f}_{1}$ for heterostructures
containing Bloch domain walls $n_{\rm Bloch}$ clearly exceeds those
for conical magnetic layers and saturates for the maximum thickness of
$n_{\rm Bloch}$ considered here. The measure of induced spin
magnetisation provided by ${\bf f}\times{\bf f}^{*}$ rises faster and
saturates earlier in heterostructures containing Bloch domain walls
compared to conical magnetic layers.

\subsection{Heterostructure including N\'{e}el domain walls}
\label{Sec3_2}

Now we focus on results obtained for heterostructures containing
N\'{e}el domain walls of increasing thickness $n_{\rm N\acute{e}el}$
to either side of the ferromagnetic middle layer. They are oriented
such that magnetic moments of the N\'{e}el domain wall align parallel
to the ferromagnetic middle layers at the interface (\fref{Fig1}(d))
and are oriented in the $yz$ plane.

Starting the discussion again with the unequal-spin spin-triplet
pairing correlations ${\bf f}_{0}$ shown for N\'{e}el domain walls of
varying thickness $n_{\rm N\acute{e}el}$ in the lower panels of
\fref{Fig2} one notices that they are nearly indistinguishable to the
respective contributions from the Bloch domain walls shown in the
middle panels of \fref{Fig2}. It seems that the orientation of the
noncollinear magnetic moments with respect to the interface plane does
not influence the induced unequal-spin spin-triplet pairing
correlations.

This changes drastically when looking at the equal-spin spin-triplet
pairing correlations where we find no contribution stemming from
heterostructures containing N\'{e}el domain walls (within the
particular setup investigated in the present work). This phenomenon
has been discussed recently by Alidoust \textit{et
  al.}~\cite{Alidoust_PRB81_014512} based on the properties of the
triplet anomalous Green's function {\bf f} used in their approach. It
turns out that for N\'{e}el domain walls to successively induce
equal-spin spin-triplet pairing correlations ${\bf f}_{1}$ the
respective noncollinear magnetic structure has to occur in the plane
parallel to the interface. In the present work the noncollinearity of
the N\'{e}el domain wall is oriented perpendicular to the interface
plane and hence does not contribute towards equal-spin spin-triplet
pairing correlations.

Focusing now at the properties of ${\bf f}\times{\bf f}^{*}$ shown in
\fref{Fig4}, one notices again that the contributions from
heterostructures containing Bloch $n_{\rm Bloch}$ (middle panels) and
N\'{e}el $n_{\rm N\acute{e}el}$ domain walls (lower panels) are again
indistinguishable. The spin magnetisation is only present in the
interface regions and does not leak into the SC regions of the
heterostructures. Again, the overall rise in spin magnetisation is
stronger for the N\'{e}el domain walls and saturates earlier than in
the conical magnetic layers (upper panels). The latter also show
stronger oscillations depending on the thickness $n_{\rm CM}$ compared
to the N\'{e}el domain wall thickness $n_{\rm N\acute{e}el}$.

In summary, for the heterostructures containing N\'{e}el domain walls
we find indistinguishable results for the unequal-spin spin-triplet
pairing correlations ${\bf f}_{0}$ and the measure of the spin
magnetisation ${\bf f}\times{\bf f}^{*}$ compared to heterostructures
containing Bloch domain walls as discussed in \sref{Sec3_1}. However,
for the setup chosen for the N\'{e}el domain walls we find no induced
equal-spin spin-triplet pairing correlations ${\bf f}_{1}$ at all.

\section{Summary and Outlook}
\label{SummaryAndOutlook}

Here we presented a detailed analysis of spin-triplet pairing
correlations occurring in SC/FM/SC heterostructures incorporating
different types of noncollinear interface regions of varying
thicknesses. Particular interfaces included conical magnetic layers,
and Bloch and N\'{e}el domain walls. It has been shown that for the
unequal-spin spin-triplet pairing correlations ${\bf f}_{0}$ both
Bloch and N\'{e}el domain walls exceed the efficiency provided by
conical magnetic layers at the interface. For the equal-spin
spin-triplet pairing correlations ${\bf f}_{1}$ only the Bloch domain
walls show considerable improvement over the conical magnetic layer
setup, whereas the specific orientation of the N\'{e}el domain wall
(noncollinear magnetic moments perpendicular to the interface plane)
prevented the occurrence of equal-spin spin-triplet pairing
correlations in that case. Different orientations of N\'{e}el domain
walls with respect to the interface plane deserve further
investigations. Finally, the dependence of the spin magnetisation
${\bf f}\times{\bf f}^{*}$ on the thickness of both, Bloch and
N\'{e}el domain walls, appears to be smoother compared to the conical
magnetic layer setup, but in all cases this stays restricted to the
interface region alone.

\ack This work has been financially supported by the EPSRC
(EP/I037598/1) and made use of computational resources of the
University of Bristol.

\end{document}